# Magnetic noise induced by dc current in a micron-size magnetic wire

A. Yamaguchi[1,2], K. Motoi[1] and H. Miyajima[1]

[1]Department of Physics, Keio University, Hiyoshi, Yokohama, Kanagawa 223-8522, Japan

[2]PRESTO, JST, Honcho, Kawaguchi, Saitama 332-0012, Japan

Present address: (A. Y) National Institute of Advanced Industrial Science and Technology, Tsukuba 305-8568, Japan

**[Abstract]**

The magnetic noise spectra induced by direct-current (dc) current flowing through a micron-scale ferromagnetic wire have been investigated. We have observed the noise spectra with a resonance frequency. Under the application of the magnetic field in the plane, the magnetic field dependences of the resonance frequency and amplitude were well interpreted by the analytical calculation based on the stochastic model. The noise spectra are attributable to the resistance oscillation reflecting the uniform magnetization precession which is produced by the Joule heating due to the dc current.

Recent interest in the magnetization dynamics in nano-scale magnet is attributed both to the fundamental physics and to the technological potential of new spintronics devices such as magnetic random access memories. Particularly the micron- and nano-scale ferromagnetic wires are suitable to investigate the spintronics associated with spin-polarized current and magnetization dynamics in a confined structure.[1] The magnetic excitation in the confined magnets is one of the main subjects in the development of the spintronics. It is well known that the spin-polarized current flowing across the ferromagnetic multilayers exerts the spin transfer torque on the magnetic moments in the layers.[2-6] The spin transfer effect also occurs in ferromagnetic films and wires with a twisted magnetic structure such as a domain wall (DW).[7-15]

The spin torque diode effect[16,17] and spin rectifying effect[18-23] have been attracted much attention because of the engineering applications and fundamental studies. These effects are produced when the radio-frequency (rf) current flows in magnetic layered films and nano-scale magnets such as a wire and disk, and they result from the magnetoresistance oscillation due to ferromagnetic resonance (FMR) generated by the spin-transfer effect[16,17] and rf magnetic fields.[18-21] In previous papers[18,21-23], we showed that the effects provide not only highly sensitive detection of the spin dynamics in nano-scale magnet but also the homodyne rf detection devices.

Well, there are several kinds of noise which restricts the signal-to-noise ratio of rf devices. In particular, the high-frequency thermal fluctuation of magnetization provides a fundamental limitation in the possibility of the scale-down device size and the increase of the operating frequency, since the fluctuation increases inversely with volume of magnets.[24] The theoretical and experimental understanding of the fundamental limitation is of practical importance. Furthermore, the study of the non-equilibrium behavior of these systems will contribute to the understanding of more complicated processes, such as the thermal nucleation of domain structures and the damping process of magnetization accompanied with spin waves. These problems have been studied and can be approached through simplifications that have proved successful in the theory of the Brownian motion and other stochastic processes. The magnetic noise is expected to be induced by the magnetization precession due to the thermal fluctuation in the application of an external magnetic field.[24]

The noise induced by thermal fluctuation of magnetization, which is often called as "mag-noise", has been investigated by means of giant magnetoresistance (GMR) or tunneling magnetoresistance (TMR) measurements[25–31]. Smith and Arnett firstly measured the mag-noise in the GMR head[25], and Smith also analyzed the noise spectrum based on a linearized Landau-Lifshitz-Gilbert equation in the framework of the fluctuation-dissipation theorem.[26] Safonov and Bertram described the theoretical model based on analogies with the harmonic oscillator and a seemingly different form of the fluctuation-dissipation.[27, 28] On the other hand, Rebei *et al*. gave a microscopic understanding dealing with a magnetization slightly disturbed from equilibrium.[29, 30]

In the GMR or TMR device, the contribution of the spin transfer torque should be considered in addition to the thermal fluctuation. To understand the fundamental physics of the mag-noise, the individual contributions of the fluctuation should be distinguished. The investigation of the mag-noise in the simple system, composing of a single layered ferromagnetic wire without magnetic domain controlled by the strong shape anisotropy, opens a path to understand the individual contributions.

In this paper, we study the mag-noise spectra induced by direct-current (dc) current in a single layered ferromagnetic wire. The magnetic field dependence of the peak frequency and noise power spectral density are investigated. The experimental results are also discussed based on a theory of stochastic processes with heat magnon bath.[27, 28]

At first, we consider the measured noise signal from the fluctuations of the system in a thermal equilibrium state of temperature *T*. The noise is a result of microscopic spin fluctuations due to the interaction with thermal bath. According to Safonov and Bertran,[27] we present the magnetic noise spectrum induced by the thermal fluctuation due to the Joule heating. In order to simplify their model, we discuss the thermal fluctuations in a single-domain magnetic particle with the coordinate system as shown in Fig. 1(a), wherein the longitudinal axis and short axis of the wire is parallel to the $z$- and $y$-axes in the plane, respectively. The $x$-axis is perpendicular to the plane, corresponding to the hard-magnetic anisotropy axis. The $K_1$ and $K_2$ are the easy and hard axis anisotropy constant, respectively. The out-of-plane component of the magnetization fluctuation generates the strong dynamical demagnetizing field and exerts torque proportional to $\mathbf{m} \times \delta\mathbf{m}$ onto the magnetization, rotating $\mathbf{m}$ by the angle $\delta\phi$ in the plane. Here, assuming the angle $\delta\phi$ in the plane dominates the dynamical magnetoresistnace difference, we transform the mag-noise spectrum measured as the voltage fluctuation into the projective magnetization fluctuation to the y-axis. In this case, the mag-noise is generally given by[27]

$$S_{VV}(\omega) = \int_{-\infty}^{\infty} \langle \delta V(t) \delta V(0) \rangle e^{i\omega t} dt = C_V^2 \int_{-\infty}^{\infty} \langle \delta m_y(t) \delta m_y(0) \rangle e^{i\omega t} dt. \quad (1)$$

Here, the voltage fluctuation is described by

$$\delta V(t) = C_V \delta m_y(t), \quad (2)$$

where $C_V$ denotes the factor which associates with the voltage fluctuation of the magnetoresistance generated by the magnetic moment fluctuation and is described by

$$C_V = I_{bias} \frac{\partial R}{\partial H_{ext}} \cdot \frac{\partial H_{ext}}{\partial m_y}. \quad (3)$$

According to Safonov and Bertram,[27] the microscopic model provides the resonance frequency given by

$$\omega_0 = \gamma\sqrt{(H_{\text{ext}} + H_K^\parallel + H_K^\perp)(H_{\text{ext}} + H_K^\parallel)}, \qquad (4)$$

where $\gamma$ is the gyromagnetic ratio. Generally assuming that the present system is treated as the single-domain magnetic particle, the easy and hard anisotropic fields are given by $H_K^\parallel = \frac{2K_1}{M_S}$ and $H_K^\perp = \frac{2K_2}{M_S}$. When the static magnetic field is applied at the angle $\theta$ from the longitudinal axis of the wire in the plane, the resonance frequency is modified as the following: [21, 22]

$$\omega_0 = \gamma\sqrt{H_{K\text{eff}}^\perp H_{K\text{eff}}^\parallel}, \qquad (5)$$

where

$$H_{K\text{eff}}^\parallel = H_{\text{ext}}\cos(\theta - \psi) + M_S(N_y - N_z)\cos 2\psi \qquad (6)$$

and

$$H_{K\text{eff}}^\perp = H_{\text{ext}}\cos(\theta - \psi) + M_S\{N_x - (N_z\cos^2\psi + N_y\sin^2\psi)\}. \qquad (7)$$

Here, $M_S$ is the saturation magnetization and $N_\alpha$ ($\alpha = x, y, z$) are the demagnetizing factors in the Cartesian coordinate system $(x, y, z)$. $\psi$ is the angle between the longitudinal axis of the wire and the effective magnetic moment vector as shown in Fig. 1(a).

Including a damping process, the mag-noise power spectral density of the complex amplitude fluctuations is theoretically given by[27]

$$S_{VV}(\omega) = \frac{\gamma k_B T}{M_S V_0}\left(\frac{H_{K\text{eff}}^\perp}{H_{K\text{eff}}^\parallel}\right)^{\frac{1}{2}}\frac{\eta}{\omega_0}F(\omega), \qquad (8a)$$

$$F(\omega) = \frac{1}{(\omega_0 - \omega)^2 + \eta^2} + \frac{1}{(\omega_0 + \omega)^2 + \eta^2}, \qquad (8b)$$

where $\eta$ and $V_0$ denote the effective damping factor and the sample volume, respectively.

The equation (8) indicates that the mag-noise is enhanced around the resonance frequency. It also shows that the amplitude of the noise power spectrum density is proportional to the system temperature $T$ and becomes suppressed with the external magnetic field.

The soft-ferromagnetic $Fe_{19}Ni_{81}$ (Permalloy: Py) wire system was fabricated by means of high-resolution electron beam lithography with a standard lift-off technique.[18, 21, 22] The Py wire with thickness of 20 nm and width of 2.2 μm was prepared onto a polished MgO (100) substrate. As shown in Fig. 1(b), the Py wire is placed within the center strip line of the coplanar waveguide (CPW) comprising Au conductive strip. The thickness of the Au electrode is 100 nm. The external magnetic field $H_{ext}$ is applied in the substrate plane along the angle $\theta$ from the longitudinal axis of the Py wire (see left side in Fig.1(b)).

The schematic circuit diagram is shown in Fig. 1(b). The Py wire of the dc resistance 48 Ω is connected with the microwave probes. The dc current is injected through a bias-tee with rf current density $J_{rf} = 4.5 \times 10^{10}$ A/m² of 10 kHz. After the rf signal induced by the dc current modulated at 10 kHz is amplified using rf-preamplifier, it is detected by a lock-in amplifier and a spectrum analyzer.[31] All measurements are performed at room temperature.

We firstly measured the rectifying effect of the Py wire and confirmed the occurrence of FMR generated by the rf magnetic field.[18] Next, using the electrical measurement circuit shown in Fig. 1(b), we measured the noise spectrum that resulted from the magnetization precession induced by the dc current flowing through the Py wire.

Figure 2 shows the rectifying spectrum (black solid-line) and the mag-noise spectrum (red solid-line) in the application of $J_{rf} = 3.3 \times 10^{10}$ $A/m^2$ and $J_{dc} = 2.7 \times 10^{11}$ $A/m^2$ in $H_{ext}$ = 100 Oe at $\theta = 45°$, respectively. It should be noted that the resonance position of the mag-noise signal almost matches that of the rectifying spectrum, indicating that the origin of the noise signal induced by dc current is the same as that of the rectifying effect.

The magnetic field dependence of the mag-noise signal in the Py wire is shown in Fig. 3. The resonance frequency increases with increasing $H_{ext}$. Figure 4 shows the magnetic field dependence of the resonance frequency, wherein the solid line shows the calculation from Eq. (5). Here, we assumed that the magnetization vector is parallel to the external magnetic field, namely, $\theta = \psi$, considering that the coercive force is small enough to direct the magnetization parallel to the external magnetic field in the range of our experimental measurement. The fitting curve described by Eq. (5) correlates well with the experimental data under the assumption of $\theta = \psi = 40°$, $M_S \approx 1.08$ T, $N_x \approx 0.96, N_y \approx 0.04$ and $N_z \approx 0$. This is interpreted that the magnetization is almost parallel to the direction of the external magnetic field.

As shown in Fig. 3, the resonant mag-noise signal disappears when $H_{ext}$ is adequately higher than the coercivity of the wire ($H_{ext} > 400$ Oe). That is, the magnetic moment in the Py wire orders along the direction of $H_{\text{ext}}$ and the precessional angle is reduced. In the consequence, the mag-noise signal due to the magnetization precession is hardly observed in the single layered films. This experimental result can be quantitatively explained by the theoretical result described by Eq. (8).[27] The inset of Fig. 5 shows the raw noise amplitude as a function of external magnetic field. The amplitude difference $\Delta V$ between the peak and baseline of the noise spectra is proportional to the ratio $\left(H_{K\text{eff}}^{\perp}/H_{K\text{eff}}^{\parallel}\right)^{1/2}$ as indicated by Eq. (8). Therefore, using the fitting curve is given by

$$\Delta V \propto \left(\frac{H_{K\text{eff}}^{\perp}}{H_{K\text{eff}}^{\parallel}}\right)^{\frac{1}{2}} + \text{offset const.} \quad . \tag{9}$$

Before fitting, we normalized the amplitude difference dividing the amplitude at $H_{\text{ext}} = 400$ Oe since the ratio $\left(H_{K\text{eff}}^{\perp}/H_{K\text{eff}}^{\parallel}\right)^{1/2}$ reached asymptotic convergence when $H_{\text{ext}} \to \infty$. To fit the normalized data in Fig. 6 by Eq. (9), we used the values $\left(H_{K\text{eff}}^{\perp}/H_{K\text{eff}}^{\parallel}\right)^{1/2}$ obtained in the fitting of Fig. 4. Equation (9) correlates well with the normalized experimental data as shown in Fig. 5. These results are in good agreement with the theoretical approach tailored for the actual system.

Next, the dc current density $J_{\text{dc}}$ dependence of the mag-noise amplitude $\Delta V$ is studied so as to clarify the contribution of the Joule heating due to the dc current. Figure 6 (a) shows the $J_{\text{dc}}$ dependence of the mag-noise spectra in the case of $H_{\text{ext}} = 100$ Oe at $\theta = 45°$. The amplitude of the mag-noise spectra is expected to be proportional to the square of the dc current density when the mag-noise is excited by the Joule heating due to the dc current. Therefore, we plotted the amplitude as a function of dc current and square of dc current in Figs. 6(b) and 6(c), respectively. The amplitude of the mag-noise spectra is increased with increasing $J_{dc}$. It makes the mag-noise amplitude be rather proportional to the square of dc current in comparison of Fig. 6(b) with Fig. 6(c). This means that the mag-noise signal is originated in the white noise excitation via the thermal fluctuation due to the Joule heating. According to Eq. (8) [27], the mag-noise spectra due to the thermal fluctuations of the magnetization is proportional to the system temperature $T$. In order to clarify the relationship between the system temperature and applied dc current, we measure the dc current density dependence of the resistance[32, 33]. The resistance is proportional to the square of dc current density $J$ as shown in Fig. 7. According to our previous experimental work,[32, 33] the system temperature is increased by about 60 K for the dc current density of $2.7 \times 10^{11} A/m^2$. The estimated system temperature contributes the thermal fluctuation of magnetization in the wire, generating the excitation of mag-noise.

The present mag-noise analysis in the single layered ferromagnetic wire offers a clue to distinguish the individual contribution for the spin dynamics induced by the spin torques, Oersted field and thermal fluctuation due to the Joule heating.

In summary, we have studied here the dc-current-induced mag-noise spectra of the single layered $Fe_{19}Ni_{81}$ wire in a magnetic field applied in the plane to the substrate. We measured the field dependence of the mag-noise spectra. The resonance frequency correlates well with the ferromagnetic resonance frequency, and the amplitude at the resonance frequency decreases with increasing the static field strength. These experimental results suggested that the mag-noise is derived from the magnetization precession induced by the thermal fluctuation due to the Joule heating. Our experimental results are in good agreement with the stochastic theoretical solutions[26-30], and they provide not only the fundamental limitation in the possibility of the scale-down size but also the enhancement of the operating frequency in the rf device applications.

**Acknowledgements**

The present work was partly supported by MEXT Grants-in-Aid for Scientific Research in Priority Areas, JSPS Grants-in-Aid for Scientific Research and PRESTO JST.

**References**

1) B. Hillebrands and K. E. Ounadjela, Spin Dynamics in Confined Magnetic Structures (Springer, Berlin, 2003), Vols. 1 - 3.

2) J. C. Slonczewski, J. Magn. Magn. Mater. **159**, L1 (1996).

3) L. Berger, Phys. Rev. **B 54**, 9353 (1996).

4) M. Tsoi, A. G. M. Jansen, J. Bass, W. C. Chiang, M. Seck, V. Tsoi and P. Wyder, Phys. Rev. Lett. **80**, 4281 (1998).

5) E. B. Myers, D. C. Ralph, J. A. Katine, R. N. Louie and R. A. Buhrman, Science **285**, 867 (1999); J. A. Katine, F. J. Albert, R. A. Buhrman, E. B. Myers and D. C. Ralph, Phys. Rev. Lett. **84**, 4212 (2000).

6) J. Grollier, V. Cros, A. Hamzic, J. M. George, H. Jaffrès, A. Fert, G. Faini, J. Ben Youssef and H. Legall, Appl. Phys. Lett. **78**, 3663 (2001).

7) A. Yamaguchi, T. Ono, S. Nasu, K. Miyake, K. Mibu, and T. Shinjo, Phys. Rev. Lett. **92**, 077205 (2004).

8) M .Yamanouchi, D. Chiba, F. Matsukura and H. Ohno, Nature **428**, 539 (2004).

9) G. S. D. Beach, C. Knutson, C. Nistor, M. Tsoi, and J. L. Erskine, Phys. Rev. Lett. **97**, 057203 (2006).

10) M. Kläui, P. –O. Jubert, R. Allenspach, A. Bischof, J. A. C. Bland, G. Faini, U. Rüdiger, C.

A. F. Vaz, L. Vila, and C. Vouille, Phys. Rev. Lett. **95**, 026601 (2005).

11) M. Hayashi, L. Thomas, Ya. B. Bazaliy, C. Rettner, R. Moriya, X. Jiang and S. S. P. Parkin, Phys. Rev. Lett. **96**, 197207 (2006).

12) L. Berger, J. Appl. Phys. **49**, 2156 (1978); L. Berger, J. Appl. Phys. **55**, 1954 (1984).

13) G. Tatara and H. Kohno, Phys. Rev. Lett. **92**, 086601 (2004).

14) S. Zhang and Z. Li, Phys. Rev. Lett. **93**, 127204 (2004).

15) A. Thiaville, Y. Nakatani, J. Miltat, and Y. Suzuki, Europhys. Lett. **69**, 990 (2005).

16) A. A. Tulapurkar, Y. Suzuki, A. Fukushima, H. Kubota, H. Maehara, K. Tsunekawa, D. D. Djayaprawira, N. Watanabe and S. Yuasa, Nature **438**, 339 (2005).

17) J. C. Sankey, P. M. Braganca, A. G. F. Garcia, I. N. Krivorotov, R. A. Buhrmann, and D. C. Ralph, Phys. Rev. Lett. **96**, 227601 (2006).

18) A. Yamaguchi, H. Miyajima, T. Ono, Y. Suzuki, and S. Yuasa, Appl. Phys. Lett. **90**, 182507 (2007); A. Yamaguchi, H. Miyajima, S. Kasai, and T. Ono, *ibid*. **90**, 212505 (2007); A. Yamaguchi, H. Miyajima, T. Ono, Y. Suzuki, and S. Yuasa, *ibid.* **91**, 132509 (2007).

19) M. V. Costache, M. Sladkov, C. H. van der Wal, and B. J. van Wees, Appl. Phys. Lett. **89**, 192506 (2006): M. V. Costache, S. M. Watts, M. Sladkov, C. H. van der Wal, and B. J. van Wees, *ibid.* **89**, 232115 (2006).

20) Y. S. Gui, N. Mecking, X. Zhou, Gwyn Williams, and C. –M. Hu, Phys. Rev. Lett. **98**,

107602 (2007); N. Mecking, Y. S. Gui, and C. M. Hu, Phys. Rev. **B 76**, 224430 (2007).

21) A. Yamaguchi, K. Motoi, A. Hirohata, H. Miyajima, Y. Miyashita, and Y. Sanada, Phys. Rev. B **78**, 104401 (2008).

22) A. Yamaguchi, K. Motoi, A. Hirohata and H. Miyajima, Phys. Rev. B **79**, 224409 (2009); A. Yamaguchi, K. Motoi, H. Miyajima, A. Hirohata, T. Yamaoka, T. Uchiyama and Y. Utsumi, Appl. Phys. Lett. **95**, 122506 (2009).

23) M. Goto, H. Hata, A. Yamaguchi, Y. Nakatani, T. Yamaoka and Y. Nozaki, J. Appl. Phys. **109**, 07D306 (2011); M. Goto, H. Hata, A. Yamaguchi, Y. Nakatani, T. Yamaoka, Y. Nozaki and H. Miyajima, to be published in Phys. Rev. B.

24) W. F. Brown, Jr. Phys. Rev. **130**, 1677 (1963).

25) N. Smith and P. Arnett, Appl. Phys. Lett. **78**, 1448 (2001).

26) N. Smith, J. Appl. Phys. **90**, 5768 (2001).

27) V. Safonov and N. Bertram, Phys. Rev. B **65**, 172417 (2002).

28) H. N. Bertram, Z. Jin and V. L. Safonov, IEEE Trans. Magn. **38**, 38 (2002).

29) A. Rebei and G. J. Parker, Phys. Rev. B **67**, 104434 (2003).

30) A. Rebei, M. Simionato and G. J. Parker, Phys. Rev. B **69**, 134412 (2004).

31) T. Seki, H. Tomita, T. Shinjo and Y. Suzuki, Appl. Phys. Lett. **97**, 162508 (2010).

32) A. Yamaguchi, S. Nasu, H. Tanigawa, T. Ono, K. Miyake, K. Mibu and T. Shinjo, Appl. Phys. Lett. **86**, 012511 (2005).

33) A. Yamaguchi, A. Hirohata, T. Ono and H. Miyajima, unpublished.

**Figure caption**

**Figure 1**

(a) The corresponding model geometry and symbol definitions. (b) The schematic diagram of the magnetoresistance oscillation and optical micrograph image of $Fe_{19}Ni_{81}$ micron-scale wire and the electrodes fabricated onto a polished MgO substrate. The magnetic field is applied in the substrate plane at an angle $\theta$ from the longitudinal wire axis.

**Figure 2**

The rf current frequency dependence of the rectifying spectrum (black line) and noise spectrum in the lockin detection signal (red line) at $\theta = 45°$ under the magnetic field of 100 Oe.

**Figure 3**

The magnetic field dependence of the noise signal generated by dc current as a function of the frequency when the external field is applied at $\theta = 45°$.

**Figure 4**

The ferromagnetic resonant frequency as a function of external magnetic field. The solid-line curve shows the fitting curve described by Eq. (5).

**Figure 5**

The normalized amplitude of the noise signal is plotted as a function of external magnetic field. The solid-line curve shows the fitting curve given by Eq. (9). The inset shows the magnetic field dependence of the raw data of the noise amplitude.

**Figure 6**

(a) The dc current dependence of the detected noise spectra is plotted as a function of frequency. The amplitude of the mag-noise is plotted as a function of dc current and [dc current]$^2$ in (b) and (c), respectively.

**Figure 7**

The dc resistance of the system as a function of the dc current density.

(a)

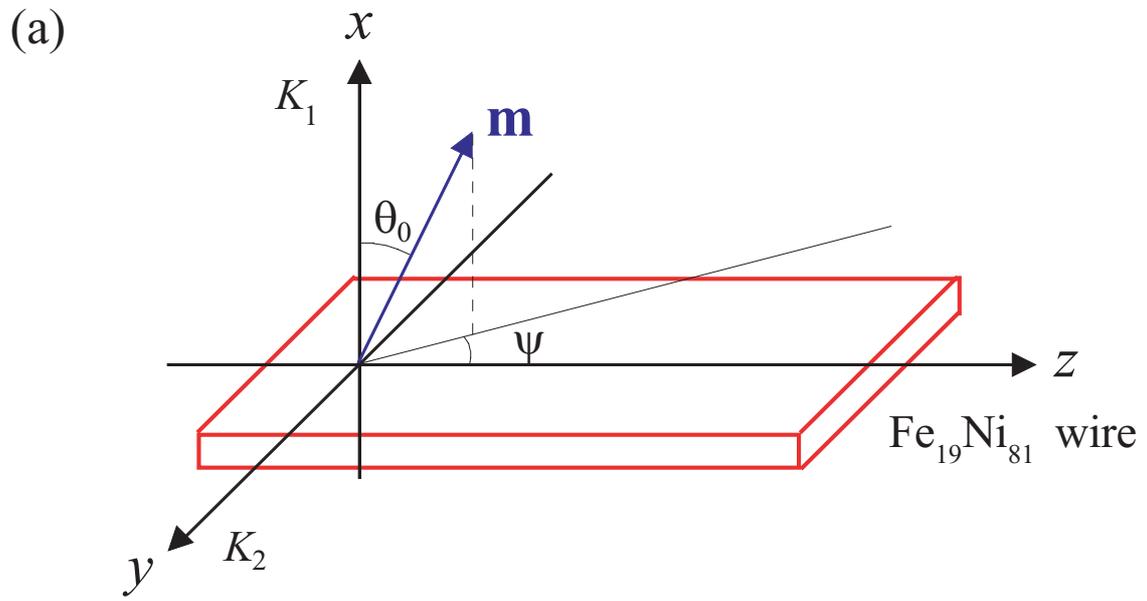

(b)

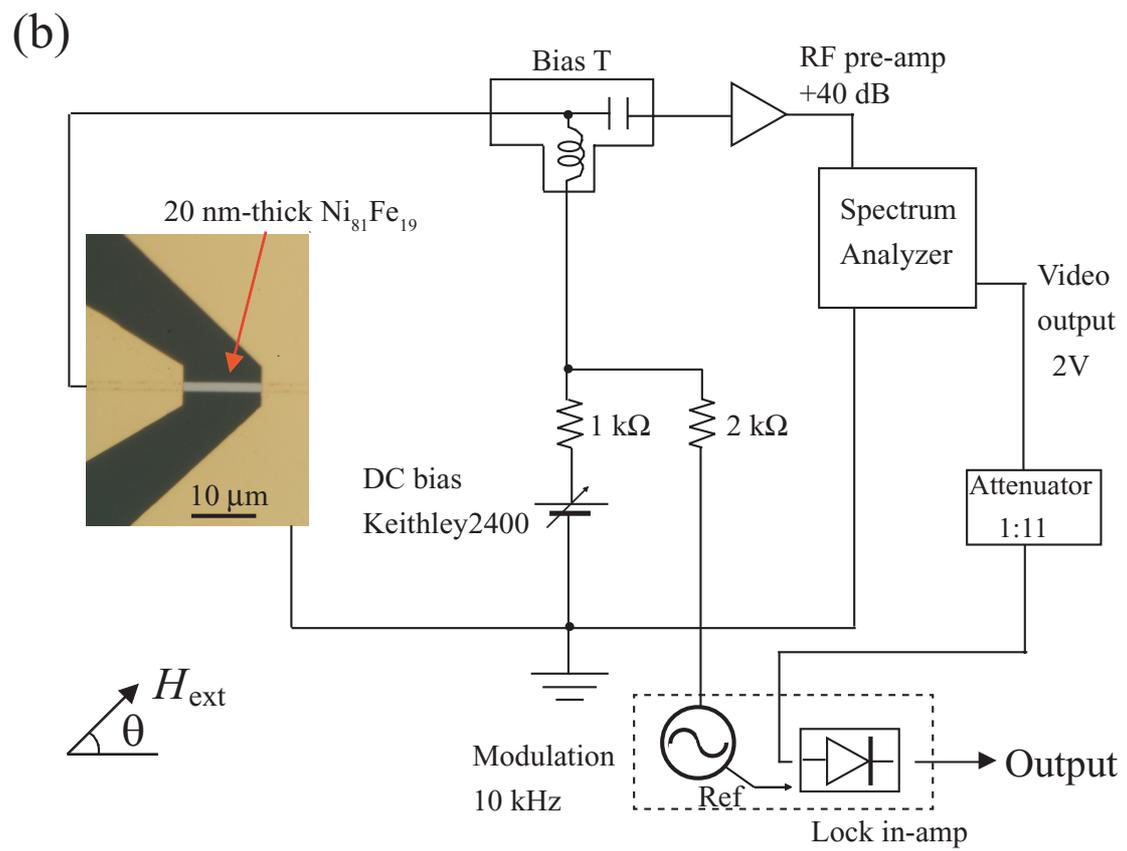

Figure 1

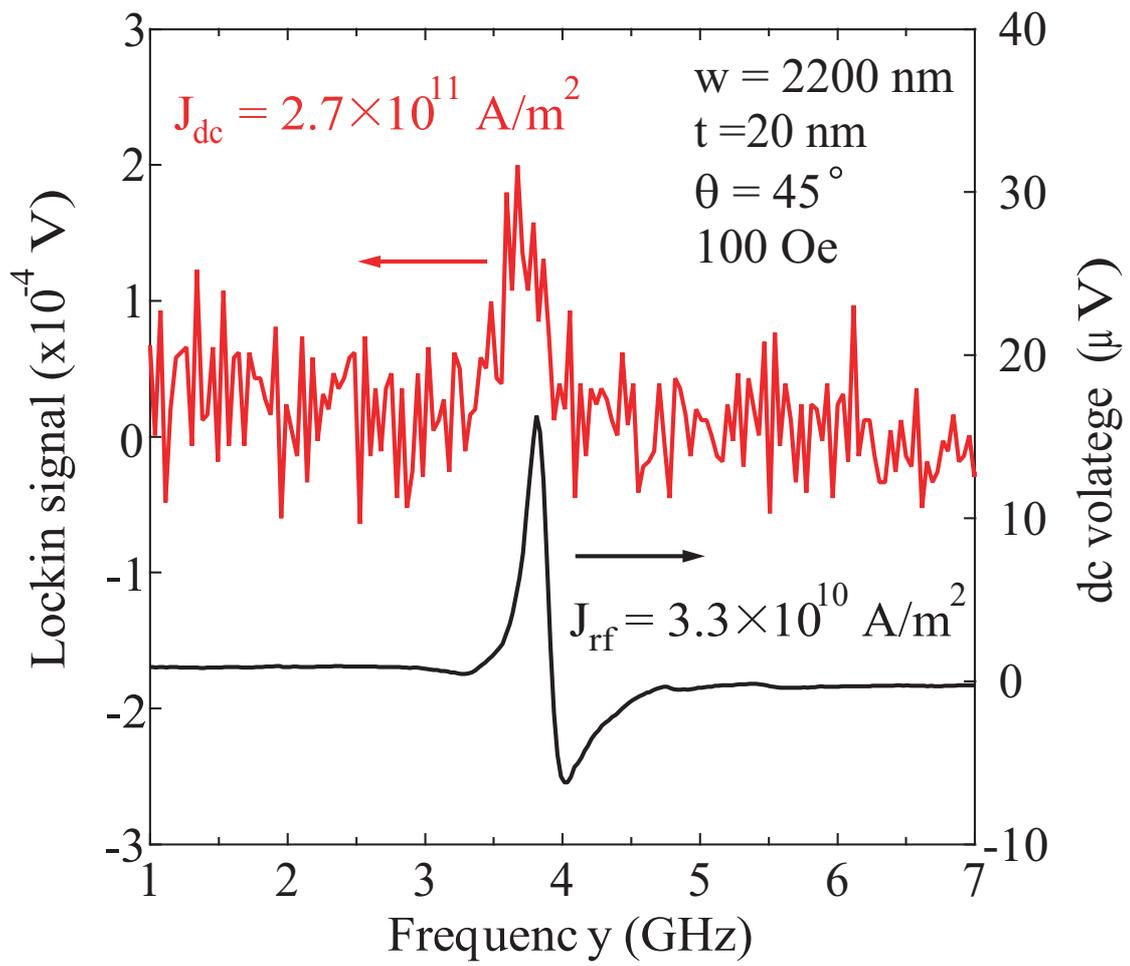

Figure 2

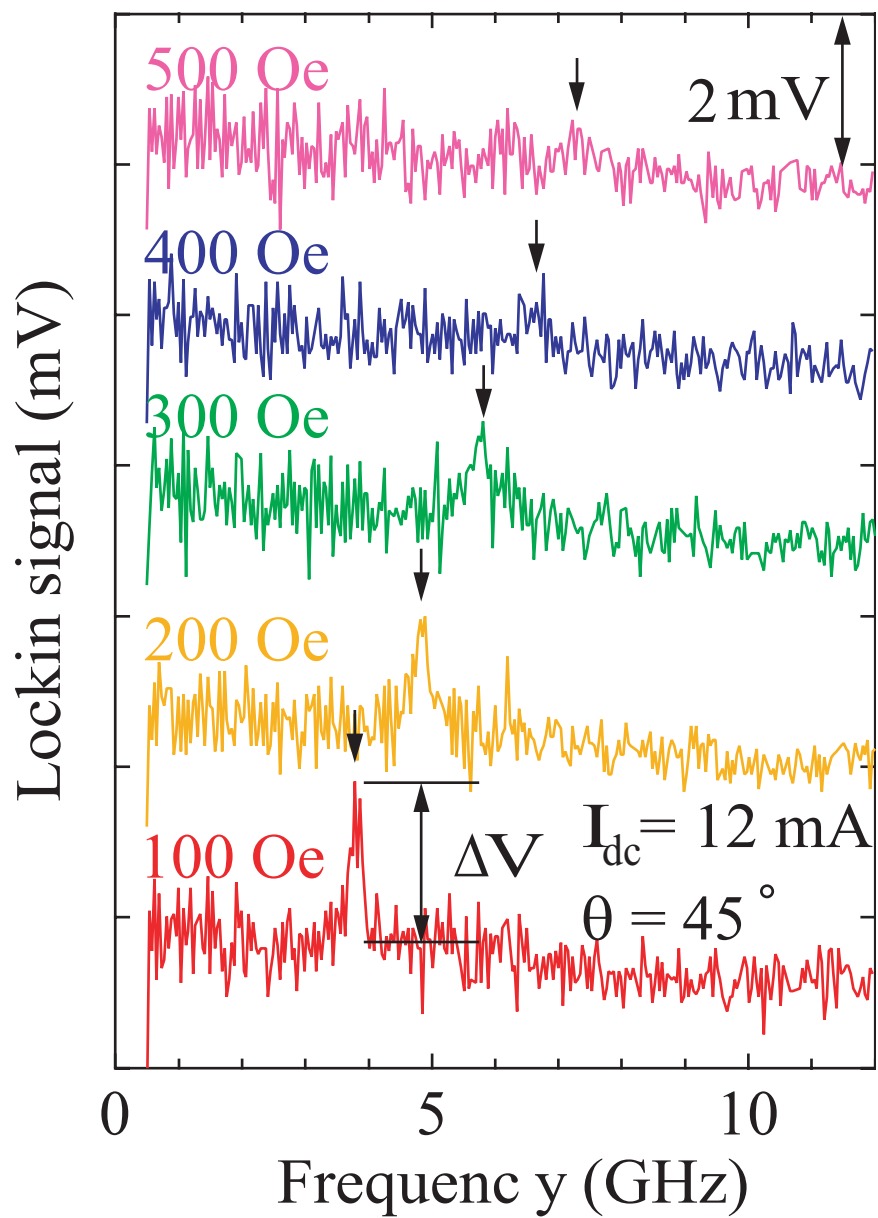

Figure 3

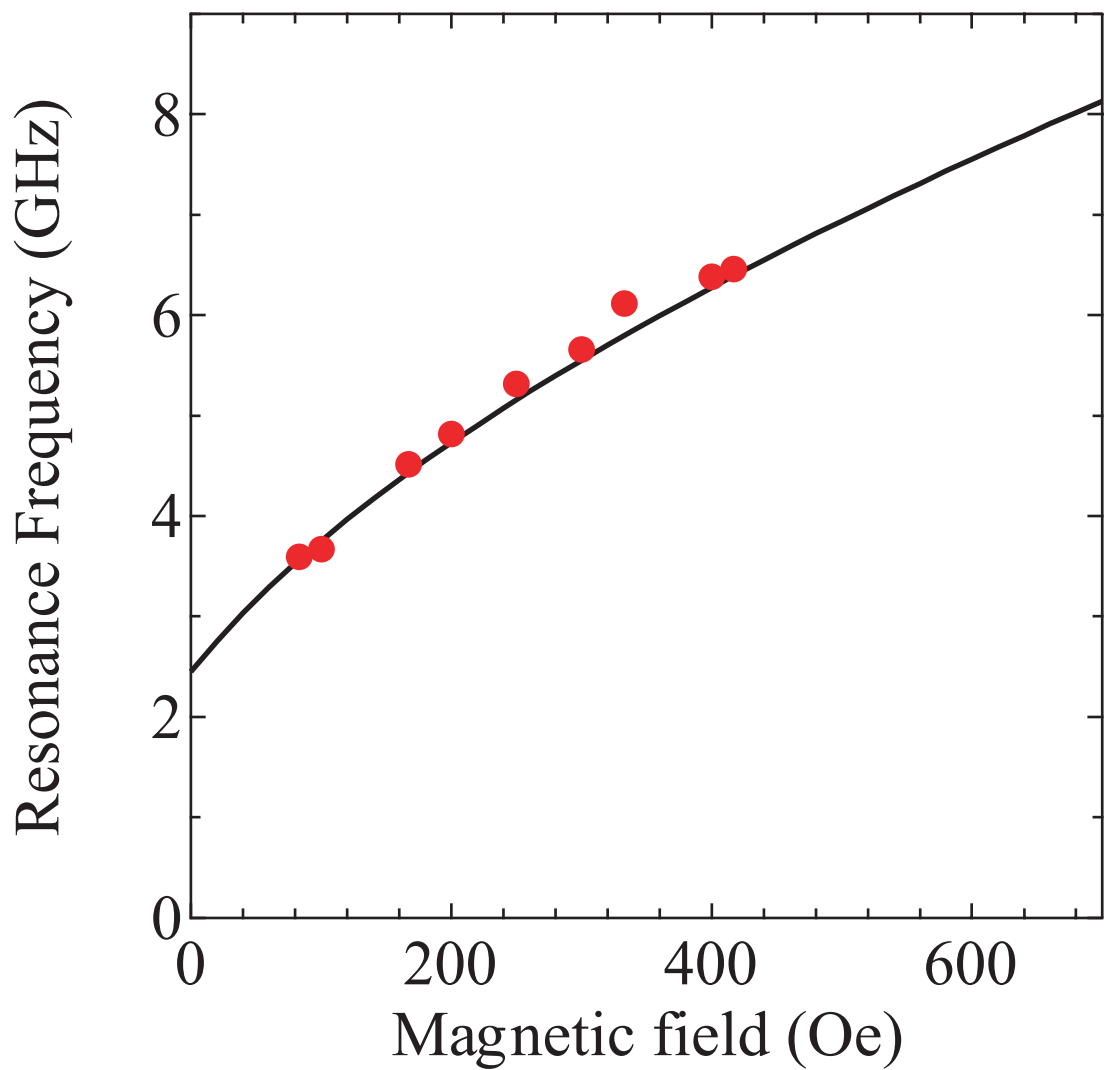

Figure 4

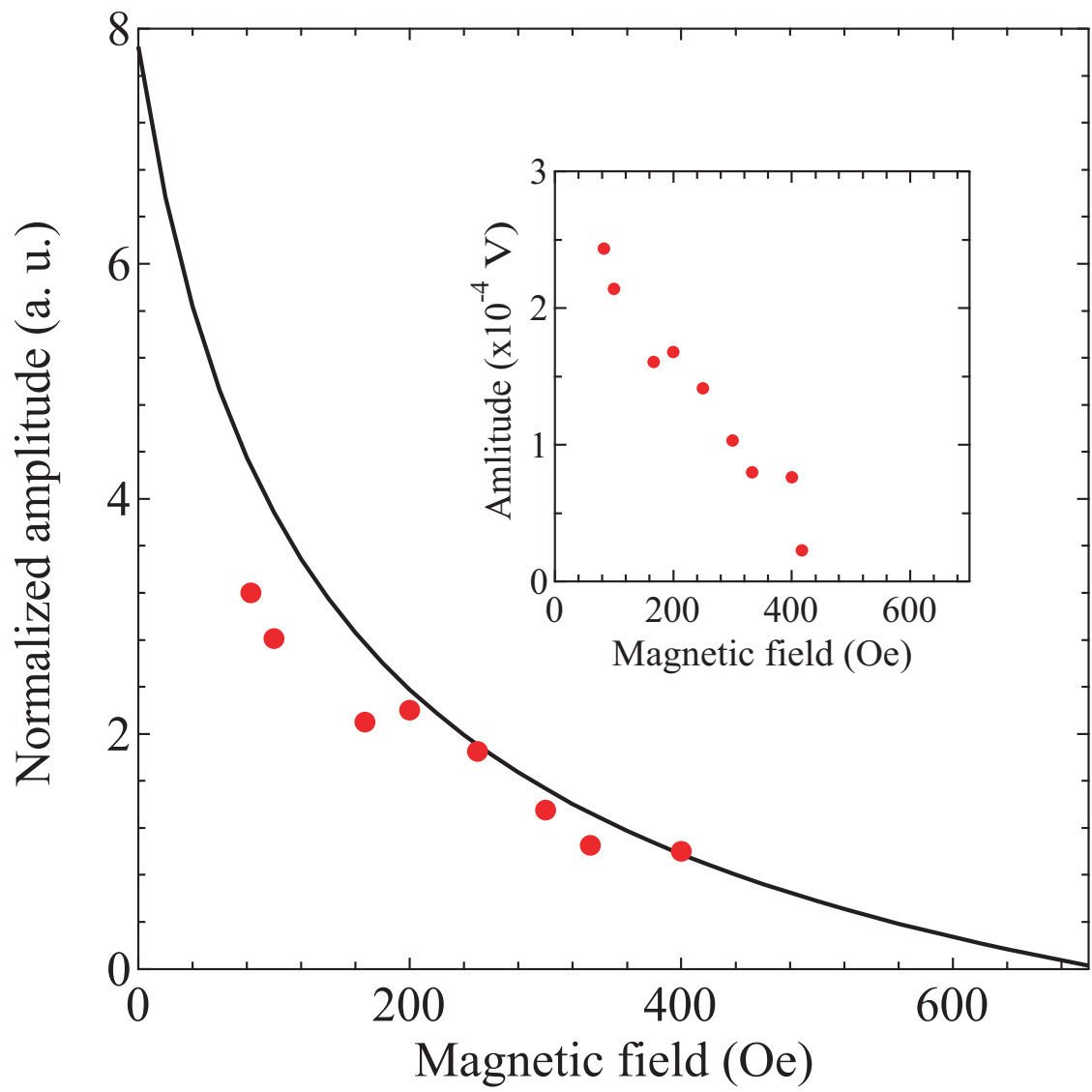

Figure 5

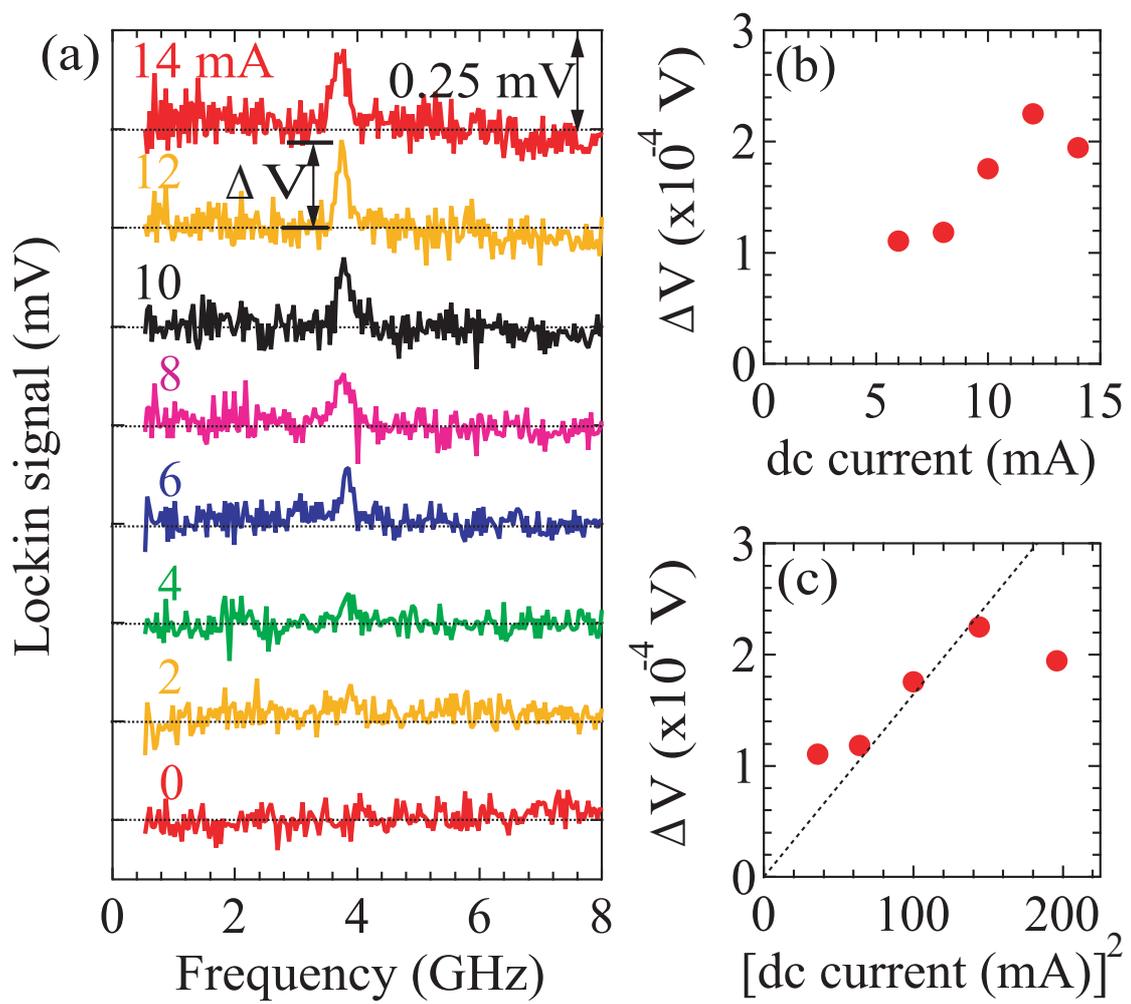

Figure 6

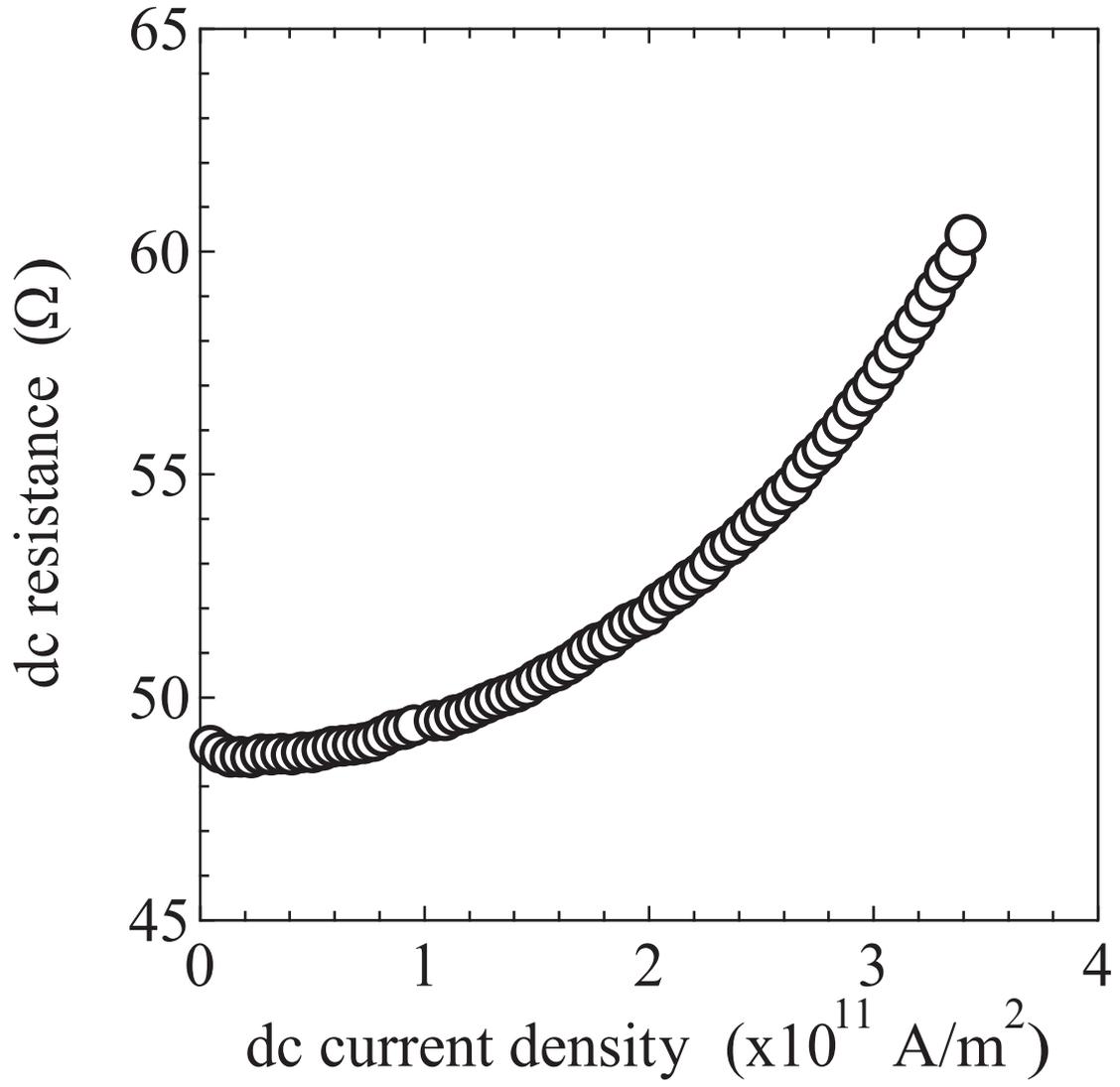

Figure 7